\documentclass[12pt]{scrartcl}

\usepackage{graphicx}
\usepackage{upgreek}
\usepackage[hyphens]{url}
\usepackage[hidelinks]{hyperref} 

\usepackage{natbib}
\usepackage{helvet}

\usepackage{url}
\usepackage[labelfont=bf, format=plain]{caption}
\usepackage{pdfpages}

\begin{document}
\date{}

\title{An automated workflow for parallel processing of large multiview SPIM recordings}

\author{Christopher Schmied\,$^{1}$, Peter Steinbach\,$^{1}$,Tobias Pietzsch\,$^{1}$, \\ Stephan Preibisch\,$^{1,2,3}$ and Pavel Tomancak\,$^1$}
\maketitle
\newcommand{\address{
\noindent
$^{1}$Max Planck Institute of Molecular Cell Biology and Genetics, Dresden, Germany \\
$^{2}$HHMI Janelia Research Campus, Ashburn, Virginia, USA \\
$^{3}$Max Delbr\"uck Center for Molecular Medicine, Berlin Institute for Medical Systems Biology, Berlin, Germany
}}
\vfill
Correspondence should be addressed to tomancak@mpi-cbg.de
\pagebreak
\begin{abstract}

\section*{Abstract:} Multiview light sheet fluorescence microscopy (LSFM) allows to image developing organisms in 3D at unprecedented temporal resolution over long periods of time. The resulting massive amounts of raw image data requires extensive processing interactively via dedicated graphical user interface (GUI) applications. The consecutive processing steps can be easily automated and the individual time points can be processed independently, which lends itself to trivial parallelization on a high performance computing (HPC) cluster. Here we introduce an automated workflow for processing large multiview, multi-channel, multi-illumination time-lapse LSFM data on a single workstation or in parallel on a HPC cluster. The pipeline relies on {\it snakemake} to resolve dependencies among consecutive processing steps and can be easily adapted to any cluster environment for processing LSFM data in a fraction of the time required to collect it. 

\section*{Availability:}
The code is distributed free and open source under the MIT license \href{http://opensource.org/licenses/MIT}{http://\\opensource.org/licenses/MIT}.
The source code can be downloaded from github: \href{https://github.com/mpicbg-scicomp/snakemake-workflows}{https://github.com/mpicbg-scicomp/snakemake-workflows}.
Documentation can be found here: \href{http://fiji.sc/Automated\_workflow\_for\_parallel\_Multiview\_Reconstruction}{http://fiji.sc/Automated\_workflow\_for\_parallel\\\_Multi-view\_Reconstruction}. 

\section*{Contact:} \href{schmied@mpi-cbg.de}{schmied@mpi-cbg.de} 
\end{abstract}

\section{Introduction}
The duration and temporal resolution of 3D fluorescent imaging of living biological specimen is limited by the amount of laser light exposure the sample can survive. LSFM alleviates this by illuminating only the imaged plane thus reducing photo damage dramatically. Additionally LSFMs achieve fast acquisition rates due to sensitive wide-field detectors and in Selective Plane Illumination Microscopy (SPIM), sample rotation enables complete coverage of large, non-transparent specimen. Taken together, LSFMs allow imaging of developing organisms {\it in toto} at single cell resolution with unprecedented temporal resolution over long periods of time (\citealp{Huisken04, Keller08}).

This powerful technology produces massive, terabyte size datasets that need computationally expensive and time-consuming processing before analysis. Existing software solutions implemented in Fiji (\citealp{Preibisch10, Preibisch14, PreibischUnpublished, Schmied14}) or in ZEISS ZEN black are performing chained processing steps on a single computer and require user inputs via a GUI. As the spatial and temporal resolution of the light sheet data increase, such approaches become inconvenient since processing can take days. 

In controlled experiments, SPIM image processing is robust enough to be automated and key steps are independent from time point to time point. HPC is inherently designed for such time consuming and embarrassingly parallel tasks that require no user interaction. Therefore, we developed an automated workflow with minimum user interaction that is easily scalable to multiple datasets or time points on a cluster. In combination with the appropriate computing resources it enables for the first time processing of SPIM data that is faster than the total acquisition time required for collecting the raw images.

\section{Processing workflow}
\begin{figure}[!tpb]
\centerline{\includegraphics{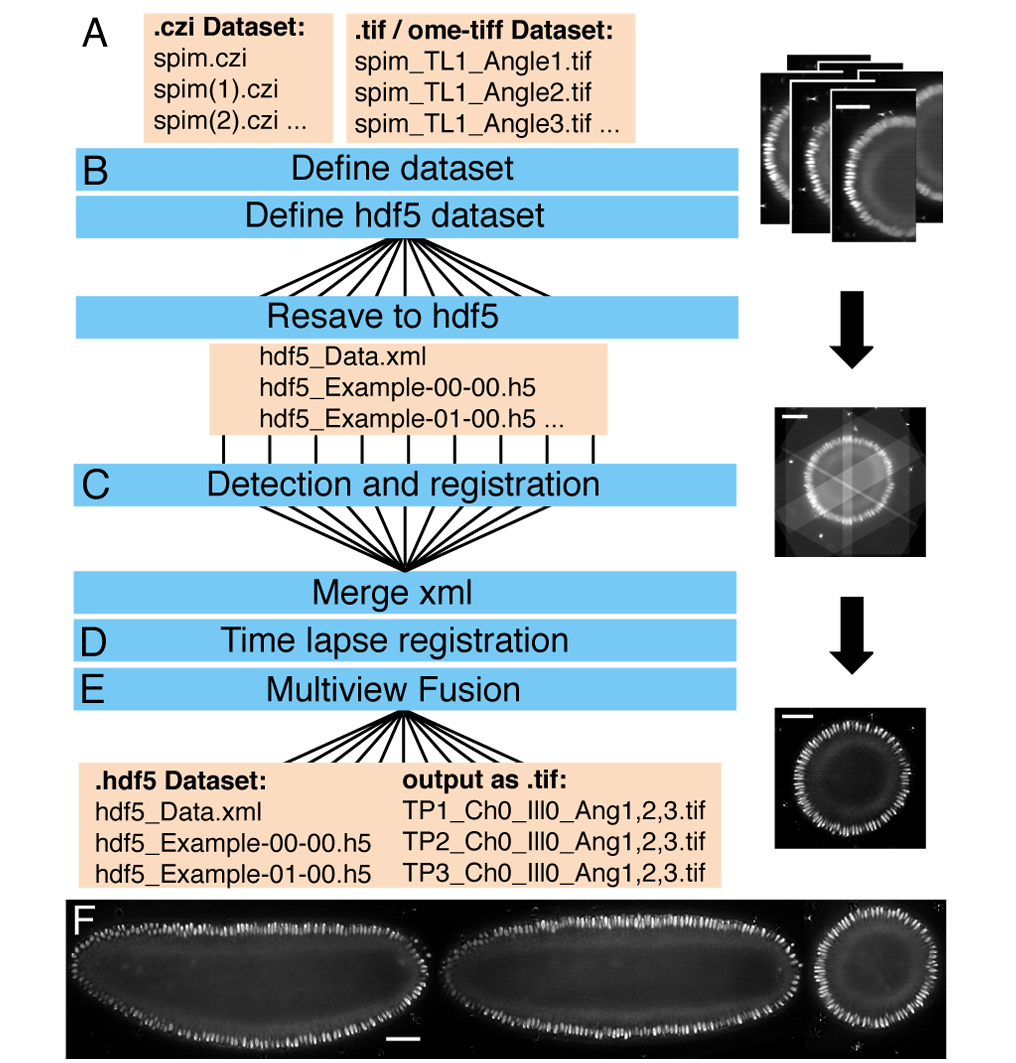}}\

\caption{Automated workflow for Multiview processing.  Workflow for SPIM image processing (A-E) using parallelization (B, C and E). Shown on the right yz--slices in the BigDataViewer of a {\it Drosophila} embryo expressing histone H2Av-mRFPruby raw (A) registered (C) and deconvolved (E). Results of deconvolution with xy--, xz-- and xz--slices through the fused volume of the same embryo (F). Scale bars represent 50 $\upmu$m.}\label{fig:01}

\end{figure}

The Fiji SPIM processing pipeline uses Hierarchical Data Format (HDF5) as data container for the originally generated TIFF or CZI files by custom made (\citealp{Pitrone13}) or commercial SPIM microscopes (Fig. \ref{fig:01}A,B). Following format conversion, multiview registration aligns the different acquisition angles (views) within each time point (Fig. \ref{fig:01}C), and subsequent time-lapse registration stabilizes the recording over time (\citealp{Preibisch10}) (Fig. \ref{fig:01}D). Fusion combines the registered views of one time point into a single volume by averaging or multiview deconvolution (\citealp{Preibisch10, Preibisch14}) (Fig. \ref{fig:01}E,F). The result is a set of HDF5 files containing registered and fused multiview SPIM data that can be examined locally or remotely using the BigDataViewer (\citealp{Pietzsch15}).

All steps are implemented as plugins (\citealp{Preibisch10, Preibisch14, PreibischUnpublished, Pietzsch15}), in the open-source platform Fiji (\citealp{Schindelin10}).We use these plugins by executing them from the command line as Fiji beanshell scripts (Suppl. Fig. 1). To overcome the legacy dependency of Fiji on the GUI we encapsulate it in a {\it  virtual framebuffer} ({\it xvfb}) that simulates a monitor in the headless cluster environment (Suppl. Fig. 1). 

To map and dispatch the workflow logic to a single workstation or on a HPC cluster, we use the automated workflow engine {\it snakemake} (\citealp{Koster12}). The workflow is defined using a {\it Snakefile} containing the name, input and output file names of each of the processing steps and python code calling the {\it beanshell scripts} (Suppl. Fig. 1). Upon invocation, the {\it snakemake} rule engine resolves the dependencies between individual processing steps based on the input files required and the output files produced during the workflow. It also creates the command that fits the input/output rule description and the template command as defined in the {\it Snakefile}. Most importantly, if single tasks on individual files are discovered to be independent, they are invoked in parallel (Suppl. Fig. 2). Each instance of {\it snakemake} for one dataset is independent and thus the workflow can be applied simultaneously to multiple dataset.

The required parameters for processing are collected by the user during GUI processing of an exemplary time point and entered into a {\it.yaml} configuration file (Suppl. List 1). The workflow is executed by passing the {\it .yaml} file to {\it snakemake} on the command line (Suppl. Fig. 1). Importantly, from the user perspective the launching of the pipeline on a HPC cluster and on a local workstation appears identical and require a single command (Suppl. List 2).  If the parameters are chosen correctly and the local or HPC resources are sufficient (Suppl. Table 1 and 2) no further action from the user is necessary. 

{\it Snakemake} supports multiple back ends to perform the command dispatch: local, cluster and  {\it Distributed Resource Management Application API} ({\it DRMAA})(\citealp{Koster12}). The local back end creates a new sub-shell and calls the command(s) required. The cluster back end is a general interface to HPC batch systems based on string substitution. {\it DRMAA} specifies a system library that interfaces all common batch systems based on a generalized task model, thus multiple batch systems are supported through one interface.

\section{Results}

\begin{table}[!h]

{\begin{tabular}{llll}
 & 1 TP local & 90 TPs local & 90 TPs cluster  \\
Resave  to hdf5 & 3 & 262 & 15 \\
Detection and registration & 2.5 & 221 & 15 \\
Average fusion & 7 & 661 & 47 \\
Deconvolution (GPU) & 21 & 1874 & 740 \\
Resave output & 3 & 286 & 7 \\
Total with average fusion  &  & 23 h 56 min & 1 h 31 min \\
Total with deconvolution (GPU) & & 44 h 08 min & 13 h 10 min \\
\end{tabular}}
\caption {Processing time comparison. Time (minutes) for key processing steps that are parallelized on a cluster. Total processing time including non parallel processing steps on the example dataset using either average fusion or deconvolution.\label{Tab:01}}

\end{table}

We compared the performance of the pipeline on a 175 GB, single channel SPIM recording of a {\it Drosophila} embryo consisting of 90 time points and 5 views, processed either on a single computer or on a HPC cluster (Table~\ref{Tab:01}). The processing using average fusion takes almost precisely one day on a single powerful computer. In contrast, using the full cluster resource the dataset can be processed in 1 h 31 min, which represents a 16-fold speedup in processing. Since the time-lapse covers 23 hours of {\it Drosophila} embryonic development the processing becomes real time with respect to the acquisition. Using deconvolution on a cluster with only 4 GPUs (Suppl. Table 1) still brings a more than 3-fold speed up (Table~\ref{Tab:01}). A dataset of 2.2 TB in size with 715 time points (\citealp{Schmied14}) would take an estimated week to process on a single computer. Using this method the processing is reduced to only 15~h with typical cluster workload from other users.  

\section{Conclusion and Outlook}

The biologist`s goal is to analyse, for instance, cellular behaviour using time-lapse SPIM recordings. The steps between data acquisition and analysis are of rather technical interest. Our pipeline leverages HPC to reduce the notoriously difficult and time-consuming SPIM data processing to a single autonomous command. Future improvements of the workflow will provide greater accessibility to novice users by using the UNICORE GUI framework (\citealp{Almond99}). Ultimately, we aim for a completely unsupervised automated processing similar to grid computing practiced in fields facing similar big data challenges such as particle physics and molecular simulation (\citealp{Bird11, Gesing12})

%
%

\section*{Acknowledgement}

We thank Stephan Janosch for valuable discussions and Akanksha Jain for testing the workflow. We thank the computer services of the MPI-CBG for their great general support and specifically Oscar Gonzalez, the members of the scientific computing facility and light microscopy facility.

\section*{Funding} P.T. and C.S. were supported by the HFSP Young Investigator grant RGY0093/2012. P.T. and T.P. were supported by the European Research Council Community`s Seventh Framework Program (FP7/2007-2013) grant agreement 260746.

%
%

\setcounter{figure}{0}

\renewcommand{\figurename}{Suppl. Fig.}

\pagebreak
\section*{Supplementary data} 
\begin{figure}[!pb]
\centerline{\includegraphics{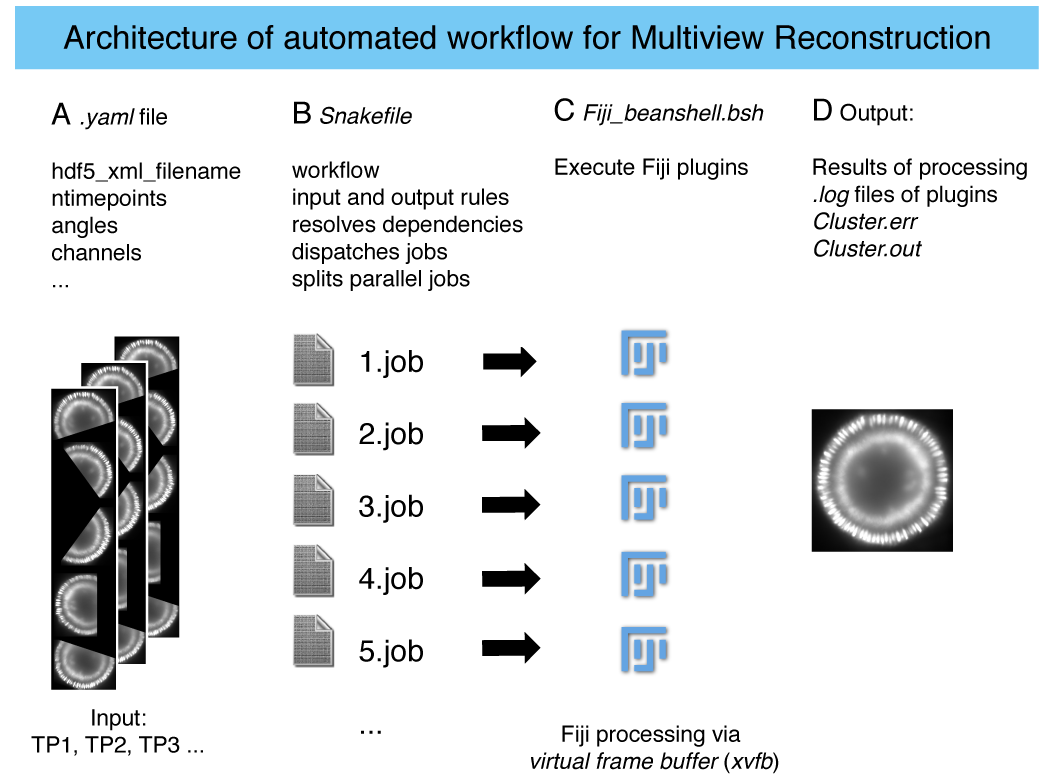}}\

\caption{Conceptual architecture for processing a multiview dataset.Time-lapse recording of Histone-YFP expression during {\it Drosophila melanogaster } embryogenesis with 5 views. The parameters are determined prior to the automated processing and stored in a .yaml configuration file (A). These parameters are passed to a Snakefile, which contains the logic of the workflow (B). Upon execution of snakemake and presence of the input files (e.g. images) snakemake dispatches the jobs which call Fiji beanshell scripts to carry out the processing using Fiji (C). The output generated by the workflow triggers the next batch of jobs once the input rules of the next step are fulfilled. Additionally, the processing writes log files and the cluster error and output files (D).}\label{Suppl. Fig.:01}

\end{figure}

\begin{figure}[!tpb]
\centerline{\includegraphics{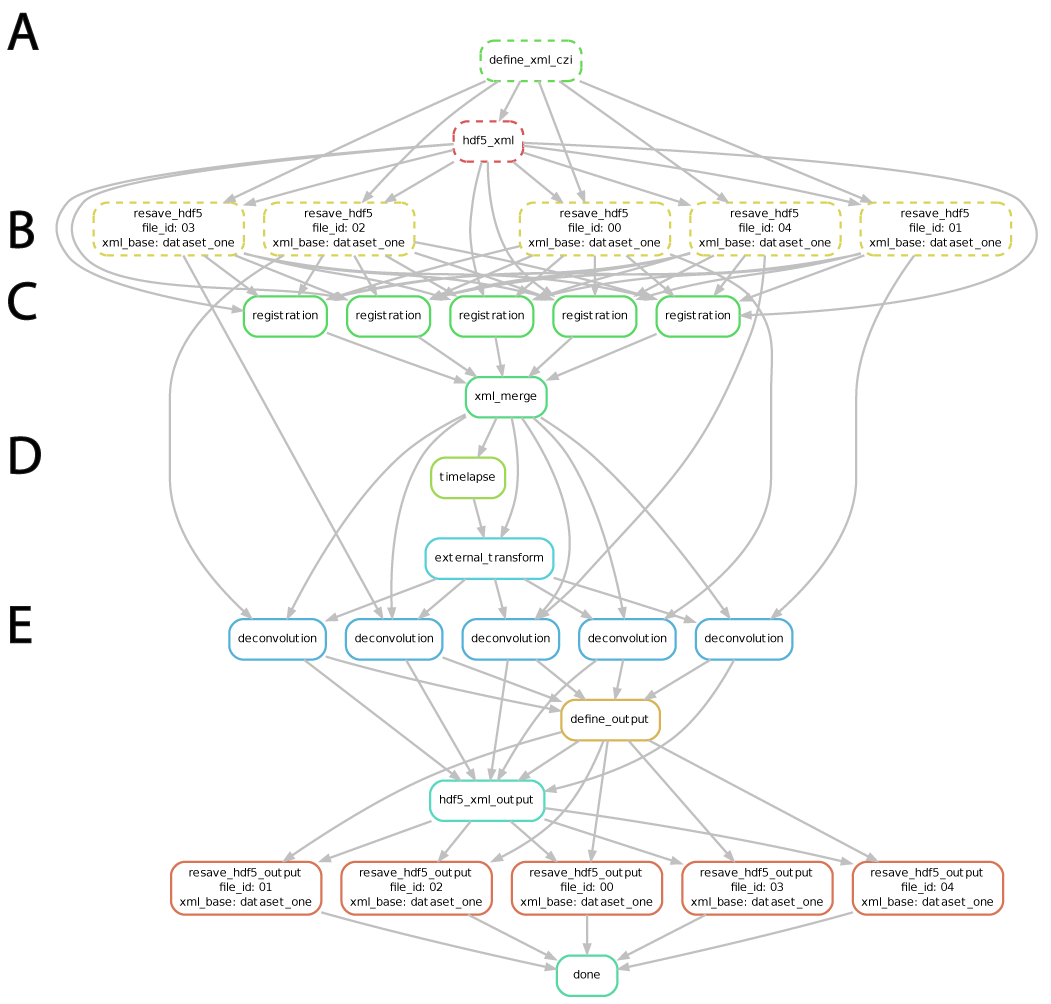}}\

\caption{Dependency graph of the snakemake workflow. Example of a directed acyclic graph (DAG) for processing a dataset with 5 time points. Snakemake resolves the file dependencies (arrows) between the different processing steps (boxes, each step with different colour). Jobs are dispatched when the input rule of the first processing step is fulfilled (A). The next batch is sent when all outputs of the processing step are created and the input rule of the next step is fulfilled (B-E). Independent tasks in the same processing step are dispatched in parallel, i.e. parallel processing of time points (B, E). }\label{Suppl. Fig.:02}

\end{figure}

\begin{figure}[!tpb]
\centerline{\includegraphics{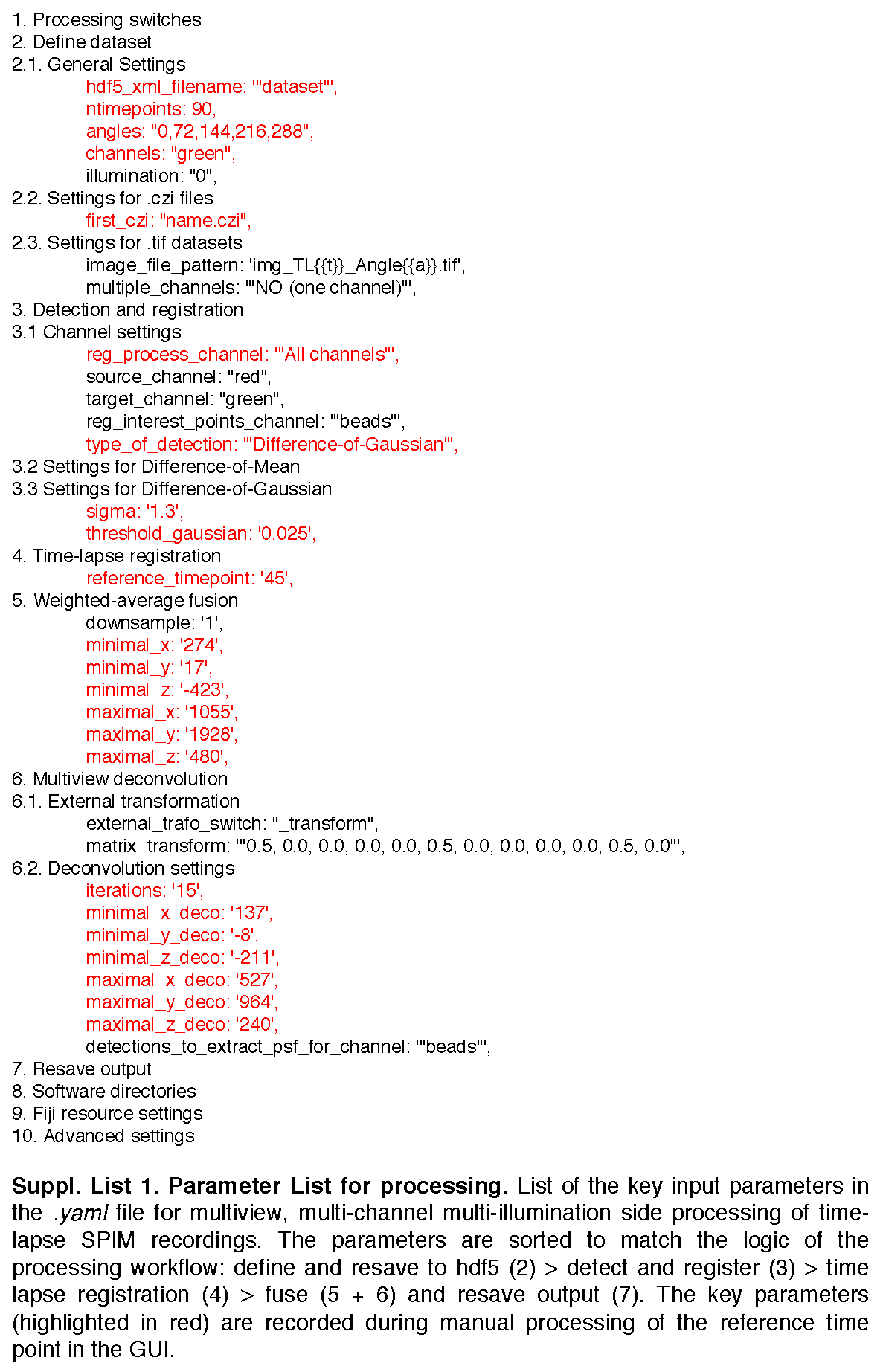}}\

\end{figure}

\begin{figure}[!tpb]
\centerline{\includegraphics{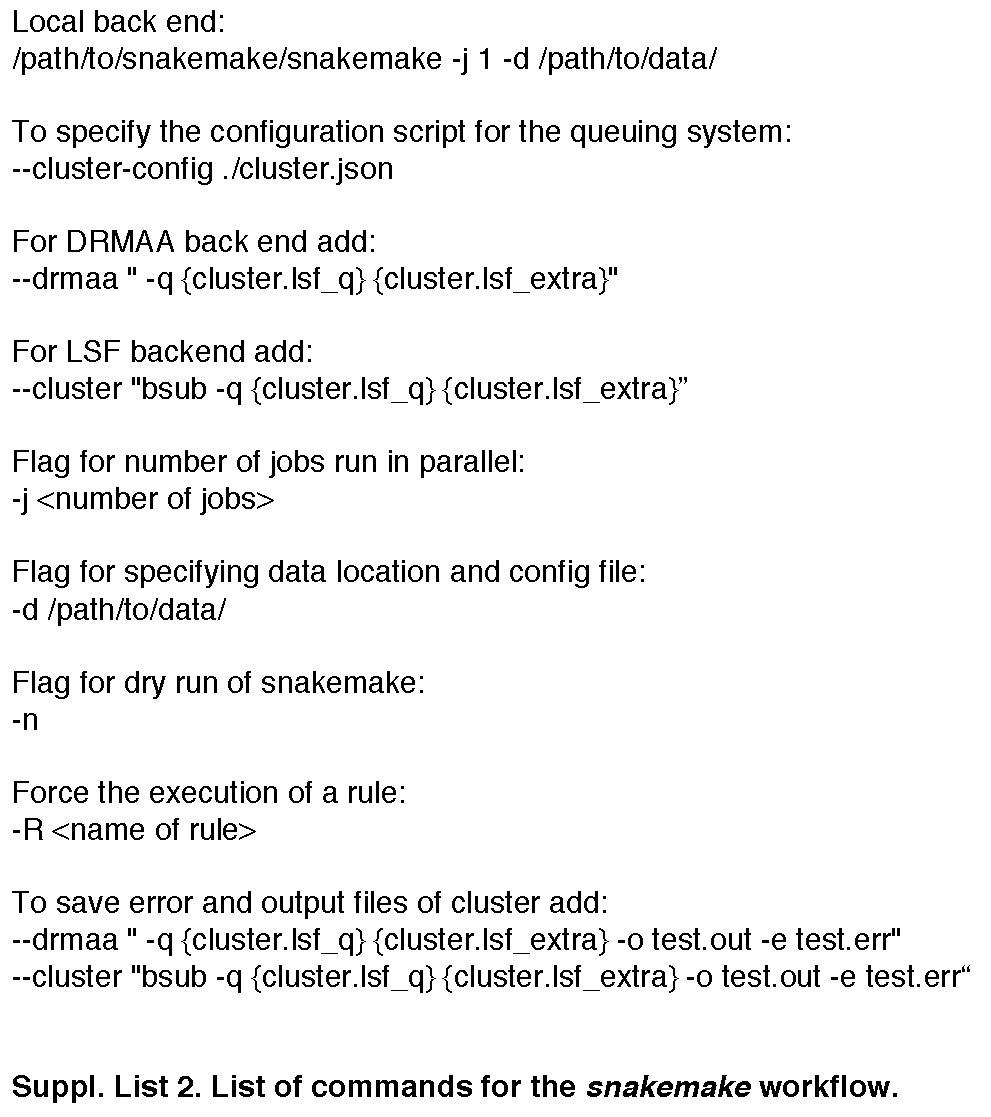}}\

\end{figure}

\begin{figure}[!tpb]
\centerline{\includegraphics{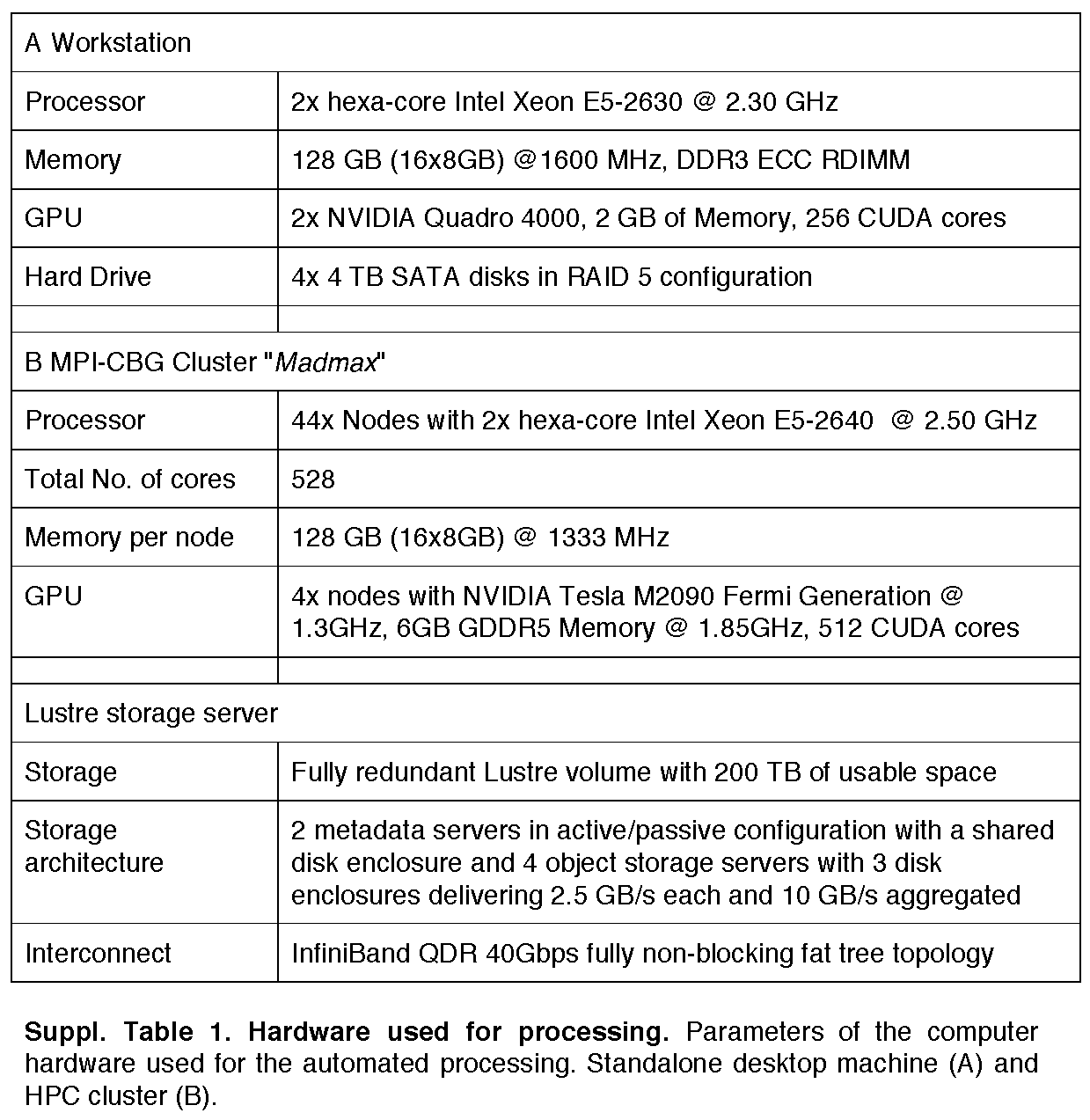}}\

\end{figure}

\begin{figure}[!tpb]
\centerline{\includegraphics{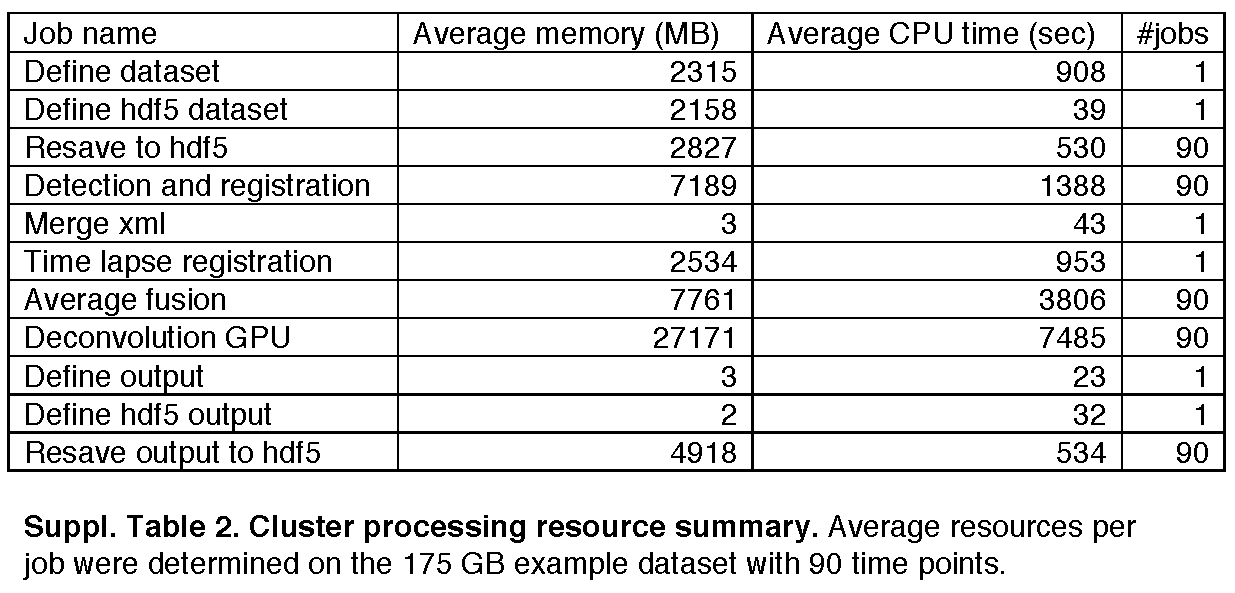}}\

\end{figure}
\end{document}